\documentclass{emulateapj}
\usepackage{apjfonts}

\newcommand{\chandra}{{\it Chandra}}

\newcommand{\rosat}{{\it ROSAT}}

\newcommand{\xmm}{{\it XMM-Newton}}
\newcommand{\einstein}{{\it Einstein}}
\newcommand{\lum}{\thinspace\hbox{$\hbox{ergs}\thinspace\hbox{s}^{-1}$}}
\newcommand{\flux}{\thinspace\hbox{$\hbox{ergs}\thinspace\hbox{cm}^{-2}\thinspace\hbox{s}^{-1}$}}
\newcommand{\hst}{{\it HST}}

\makeatletter

\begin{document}

\def\spose#1{\hbox to 0pt{#1\hss}}
\def\laeq{\mathrel{\spose{\lower 3pt\hbox{$\mathchar"218$}}
     \raise 2.0pt\hbox{$\mathchar"13C$}}}
\def\gaeq{\mathrel{\spose{\lower 3pt\hbox{$\mathchar"218$}}
     \raise 2.0pt\hbox{$\mathchar"13E$}}}

\slugcomment{Accepted for publication in ApJ}

\title{The Ultraluminous X-ray Sources near the Center of M82}
\author{A.~K.~H.~Kong\altaffilmark{1}, Y.~J.~Yang\altaffilmark{2}, P.-Y.~Hsieh\altaffilmark{2}, 
D.~S.~Y.~Mak\altaffilmark{3}, and C.~S.~J. Pun\altaffilmark{3}}

\altaffiltext{1}{Kavli Institute for Astrophysics and Space Research,
Massachusetts Institute of Technology, Cambridge, MA 02139}
\altaffiltext{2}{Institute of Astronomy and Astrophysics, Academia 
Sinica, Taipei, Taiwan}
\altaffiltext{3}{Department of Physics, University of Hong Kong, 
Pokfulam, Hong Kong}

\begin{abstract}
We report the identification of a recurrent ultraluminous X-ray source
(ULX), a highly absorbed X-ray source (possibly a background AGN), and a young supernova remnant near the center
of the starburst galaxy M82. From a series of \chandra\
observations taken from 1999 to 2005,  
we found that the transient ULX first appeared in 1999 October. The source
turned off in 2000 January,  
but later reappeared and has been active since then. The X-ray
luminosity of this source varies from below the
detection level ($\sim 2.5\times10^{38}$\lum) to its active state in between 
$\sim7\times10^{39}$ ergs s$^{-1}$ and $1.3\times10^{40}$ ergs s$^{-1}$  
(in the 0.5-10 keV energy band) 
and shows unusual spectral changes. The X-ray spectra of some
\chandra\ observations are best fitted
with an absorbed power-law model with  
photon index ranging from 1.3 to 1.7. These spectra are similar to those
of Galactic black hole binary candidates  
seen in the low/hard state except that a very hard spectrum was
seen in one of the observations. By comparing with near infrared images taken with
the {\it Hubble Space Telescope}, the ULX is found to be located  
within a young star cluster. Radio imaging indicates that it is
associated with a \ion{H}{2} region. We  
suggest that the ULX is likely to be a $> 100 M_\odot$ 
intermediate-mass black hole in the low/hard state. In  
addition to the transient ULX, we also found a highly absorbed hard
X-ray source which is likely to be an AGN and an ultraluminous  
X-ray emitting young supernova remnant which may be related to a
100-year old gamma-ray burst event, within 2 arcsec of the transient ULX.
\end{abstract}

\keywords{black hole physics --- galaxies: individual (M82) --- supernova remnants --- X-rays: binaries --- X-rays: 
galaxies}

\section{Introduction}
Ultraluminous X-ray sources (ULXs) are defined as off-nuclear X-ray
sources with isotropic luminosities  
much higher than the Eddington limit for a solar mass black hole ($L_X 
\sim 1.3\times10^{38}$ ergs s$^{-1}$).
Typical X-ray luminosities of ULXs are
 in between $10^{39}$\lum\ and $10^{41}$\lum. The physical nature of
 ULXs has been an enigma because of their high energy output. Many ULXs show strong
 variability suggesting that they are accreting compact objects.
Assuming the emission is
 isotropic, then some of the ULXs may harbor intermediate-mass black hole
 (IMBH; Colbert \& Mushotzky 1999; Makishima et al. 2000) 
 with masses of $100-10000 M_\odot$. Alternatively, ULXs may 
 simply be stellar-mass black holes. It has been suggested that ULXs 
are stellar-mass black holes with radiation pressure-dominated
(Begelman 2002) or slim (Ebisawa et al. 2003) accretion disks that
cause super-Eddington luminosities.
 Furthermore, ULXs may be
stellar-mass black holes with anisotropic X-ray emission (King et al. 2001), 
or micro-blazars which happened to be observed along the direction of
their relativistically beamed jet
(K\"ording et al. 2002). 
In addition, some ULXs may be young X-ray luminous
supernova remnants in a high-density medium, 
or hypernova remnants. Finally, some ULXs have been identified with
background AGNs through optical follow-up spectroscopy (e.g., Foschini
et al. 2002; Masetti et al. 2003).
Each of these models has difficulties to fully
explain the observations, but yet  has some supporting pieces of
evidence. Currently, we do not have a complete picture about the
physical nature of ULXs, primarily because we do not have 
dynamical mass measurements of the compact objects that power ULXs. 

In our current stellar formation and evolution theory, 
we are only able to constrain two classes of black holes, the
supermassive black holes with masses exceeding $10^{6} M_\odot$ at
the center of galaxies and
stellar-mass black holes with masses lower than 20 $M_\odot$.  
If some ULXs host IMBHs, they might provide a clue to fill in the missing
link between stellar-mass black holes and supermassive black holes. 

In this paper, we report on a recurrent transient ULX in the 
starburst galaxy M82 by using 
archival \chandra\ and {\it Hubble Space Telescope (HST)} data. The source is
located near the galactic dynamical center and is one of the 
most luminous X-ray sources within the supper-bubble region of M82  
(Matsushita et al. 2005; Matsumoto et al. 2000), close to several super-star clusters.
This source is probably the second most luminous source in M82. The 
most luminous source, M82 X--1, is a prime candidate of IMBH
and is about 5 arcsec from the transient (see Kaaret et al. 2006 and
references therein). In addition, we also study the 
physical nature of the two ULXs very close to the ultraluminous 
transient by using multi-wavelength data. 

In \S\,2 we describe the \chandra\ and \hst\ observations. We present
the data analysis and results in \S\,3. A discussion of the three ULXs
is given in \S\,4.

\section{Observations}
\subsection{\chandra}
M82 (NGC 3034) is a nearby starburst galaxy. We adopt a distance of
3.6 Mpc to M82 based on the Cepheid distance of $3.63\pm0.34$
Mpc to its close neighbor galaxy M81 (Freedman et al. 1994). 
M82 was observed
twelve times between the year of 1999 and 2005  
with \chandra.
The details of the observations are given in Table 1. 
Among these twelve observations, observations 3 and 5 were 
using the High Resolution Camera 
(HRC-I). The rest were using the Advanced CCD Imaging 
Spectrometer array (ACIS-I or ACIS-S).
We used CIAO v3.3, HEAsoft v6.2, and XSPEC v11.3 packages to perform data 
reduction and analysis. 

For ObsIDs 1411 and 380, there are two separate observations 
within the same event list. Therefore, we used a time-filter to split 
the observations and analyzed the data separately (observations 3, 5,
7, and 8).
Five of the observations (4, 6, 7, 8, and 9) are off-axis 
to reduce pile-up of M82 X--1 due to its high luminosity. In
particular, during observations 10--12, the detector 
employed a 1/8 subarray mode with a frame
time of 0.441\,s to reduce pile-up.

\begin{table}
\centering{\tiny
\caption{\chandra\ Observation Log}
\begin{tabular}{clcccc}
\hline
\hline
Index &Date & ObsId & Exposure & Instrument & Remark \\
\hline
1&$1999-09-20$ & 361 &33.7 ks & ACIS-I \\
2&$1999-09-20$ & 1302 & 15.7 ks & ACIS-I \\
3&$1999-10-28$ & $1411-1$ &36.3 ks & HRC-I \\
4&$1999-12-30$ & $378$&4.2 ks & ACIS-I  & off-axis\\
5&$2000-01-20$ & $1411-2$&17.8 ks & HRC-I \\
6&$2000-03-11$ & $379$&9.1 ks & ACIS-I  & off-axis\\ 
7&$2000-05-07$ & $380-1$&3.9 ks & ACIS-I& off-axis  \\
8&$2000-06-12$ & $380-2$&1.2 ks & ACIS-I& off-axis  \\
9&$2002-06-18$ & $2933$&18.3 ks & ACIS-S \\
10&$2005-02-04$ & $6097$&58.2 ks & ACIS-S & off-axis, 1/8 subarray\\
11&$2005-08-17$ & $5644$&75.1 ks & ACIS-S & 1/8 subarray\\
12&$2005-08-18$ & $6361$&19.2 ks & ACIS-S &1/8 subarray\\
\hline
\end{tabular}
}
\par
\medskip
\begin{minipage}{0.95\linewidth}
NOTE.--- ObsIDs 1411 and 380 have two observations merged in one event
list. We used a time filter to separate the two observations.
\end{minipage}
\end{table}

\subsection{\hst}
Since M82 has high extinction near its core where our targets are
located, we used near IR image to study the environment around the
X-ray sources.
We obtained \hst\ Near Infrared Camera and Multi-Object Spectrometer
(NICMOS) data from the Multimission Archive at STScI (MAST). M82 was 
observed on 1998 April 11 by using the NIC2 camera with the F160W
filter ($H$-band). The NIC2 camera has a resolution of $0''.075$ pixel$^{-1}$ and
a field-of-view of $19.2\times19.2$ arcsec$^2$. To cover the central 1.5
arcmin along the semi-major axis, we obtained 11 pipeline processed
NICMOS images from the MAST.
The individual images were combined to form a mosaic that were used for
analysis. To correct the absolute astrometry of the NICMOS image, we
aligned the NICMOS image with a wide-field \hst\ ACS mosaic image
(Mutchler et al. 2005). The ACS mosaic has a dimension of
$10'.24\times10'.24$ and can be obtained as High-Level Science
Products via the
MAST\footnote{http://archive.stsci.edu/prepds/m82}. The observations
and data reduction are described in Mutchler et al. (2005). We used the
F814W ($I$ band) mosaic for our analysis and corrected
the absolute astrometry by using the 2MASS
catalog. We identified 21 isolated stars in the field and matched with
the 2MASS catalog. Using IRAF tool {\it ccmap}, we corrected the
absolute astrometry of the ACS mosaic with a registration error of
$0.2$ arcsec. We then registered the NICMOS mosaic with the astrometric corrected wide-field ACS image.
With 9 isolated stars in both field-of-views of NICMOS and ACS mosaics, we corrected the astrometry
of the NICMOS image with a registration error of 0.036 arcsec.

\begin{figure*}[t]
\centering
\includegraphics[width=6.5in]{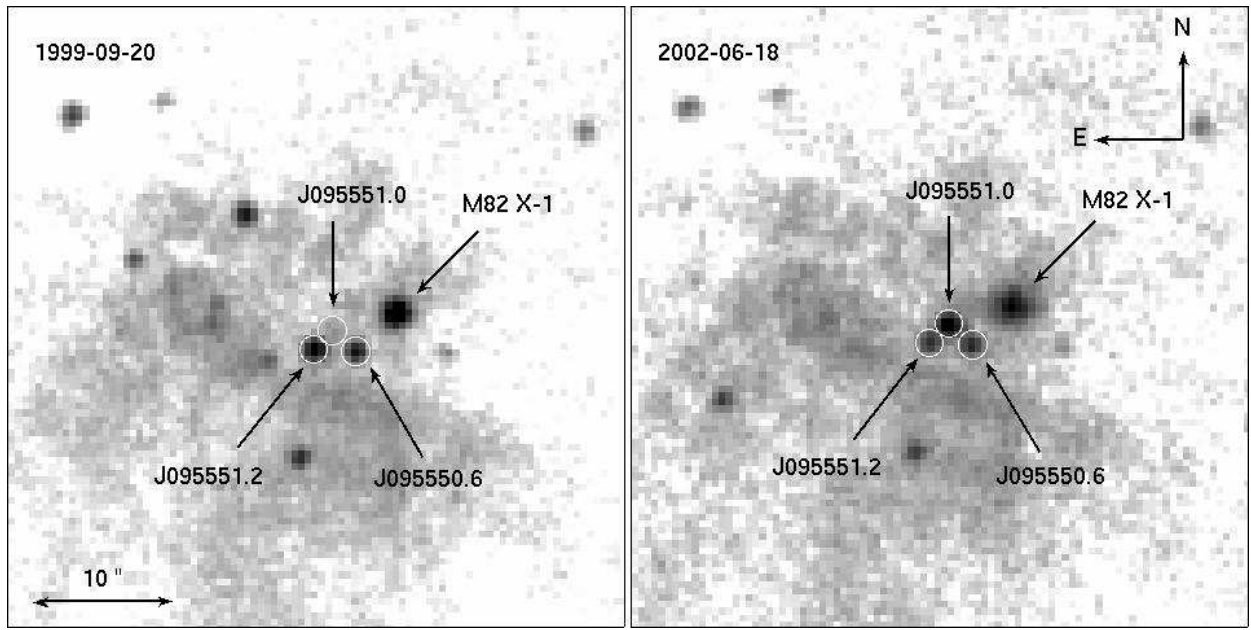}
\caption{\chandra\ 0.3--7 keV images of the central $45''\times45''$ region of
  M82 as seen on 1999 September 20 ({\it Left}; observation 1) and 2002 June 18
  ({\it Right}; observation 9). Both figures
  have the same spatial scale. The locations of
  the three ULXs discussed in this paper are marked. We also indicate
  the position of M82 X--1. The images has
  been slightly
  smoothed with a $0.5''\sigma$ Gaussian function.
}
\end{figure*}

\section{Data Analysis and Results}
\subsection{X-ray Imaging}
The brightest source in the field of M82 is M82 X--1 (CXOU
J095550.2+694047; see Fig. 1). We note that the X-ray coordinates are based on an
astrometric corrected image (see \S\,3.5). 
About 5 arcsec southeast of M82 X--1,
there is a complex of three bright X-ray sources (Matsumoto et al. 2001). By examining the
\chandra\ images, the brightest one,
CXOU J095551.0+694045 (J095551.0 hereafter; note that in Matsumoto et
al. 2001, the source is called J095551.1+694045 due to different astrometry), was clearly below the
detection limit in two observations indicating that it is a highly
variable source. Note that previous studies have mentioned the
transient behavior of this source (Matsumoto et al. 2001; Kaaret et
al. 2006). In addition to J095551.0, there are several additional
transients as shown in Figure 1; the discussion of these transients
is out of the scope of this paper.
There are two fainter X-ray sources, CXOU
J095551.2+694044 (J095551.2 hereafter) and CXOU J095550.6+694044
(J095550.6 hereafter) located at about
2 arcsec to the south of J095551.0 forming a triangle (Fig. 1). 
In contrast to J095551.0, these two sources are always active.

\begin{table*}
\centering{\footnotesize
\caption{Power-law Spectral Model for the ULXs}
\begin{tabular}{ccccccccccccccc}
\hline
\hline
Observation & \multicolumn{4}{c}{J095551.0} && \multicolumn{4}{c}{J095551.2} &&
\multicolumn{4}{c}{J095550.6}\\
\cline{2-5}  \cline{7-10} \cline{12-15}
 & $N_H$$^a$ & $\Gamma$ & $L_X$$^b$ & $\chi^2/dof$ && $N_H$$^a$ & $\Gamma$ & $L_X$$^b$
& $\chi^2/dof$ && $N_H$$^a$ & $\Gamma$ & $L_X$$^b$ & $\chi^2/dof$ \\
 & & & & & & & & & && & & & \\
\hline
1 & &&&& & $21^{+5.9}_{-2.3}$ & $2.15^{+0.95}_{-0.36}$ &
$17\pm0.7$ &1.0/81 && $2.86^{+0.45}_{-0.42}$ & $2.85^{+0.36}_{-0.33}$ & $2.4\pm0.1$& 0.9/48\\
9 & $3.63^{+0.75}_{-0.75}$ & $1.74^{+0.51}_{-0.51}$ & $7.8\pm0.2$ & 0.8/69 && $15^{+9.6}_{-3.2}$ &
$1.07^{+1.66}_{-0.61}$ & $3.2\pm0.2$ & 0.9/17 && $2.41^{+0.69}_{-0.60}$ &
$2.81^{+0.58}_{-0.52}$ & $2.0\pm0.1$& 1.1/20\\
10 & $3.19^{+0.23}_{-0.17}$ & $1.47^{+0.12}_{-0.09}$ & $12\pm0.1$ & 1.0/293\\
11 & $3.56^{+0.19}_{-0.17}$ & $1.52^{+0.09}_{-0.08}$ & $13\pm0.1$ &
1.1/331 && $13^{+2.6}_{-5.1}$ & $0.66^{+0.43}_{-0.48}$ & $3.6\pm0.09$ & 1.4/70 &&
$2.56^{+0.35}_{-0.34}$ & $2.72^{+0.31}_{-0.28}$ & $1.5\pm0.04$ & 1.5/58\\
12 & $3.34^{+0.38}_{-0.34}$ & $1.27^{+0.18}_{-0.18}$ & $11\pm0.2$ &
0.9/126 && $12^{+10.5}_{-4.6}$ & $0.61^{+1.80}_{-0.52}$ & $3.3\pm0.2$& 1.2/16
&& $3.99^{+1.30}_{-1.10}$ & $4.46^{+1.12}_{-0.93}$ & $11.6\pm0.6$ & 1.7/12\\
\hline
\end{tabular}
}
\par
\medskip
\begin{minipage}{0.95\linewidth}
NOTE.--- All quoted uncertainties are 90\% except for the luminosities
which are $1\sigma$.\\
The spectrum of J095551.0 in observation 9 suffered pile-up and a
pile-up model was applied during spectral fit.\\ 
$^a$ in units of $10^{22}$ cm$^{-2}$\\
$^b$ 0.5--10 keV unabsorbed luminosity in units of $10^{39}$ \lum\
(assuming d=3.6 Mpc). 
\end{minipage}
\vspace{3mm}
\end{table*}

\subsection{X-ray Spectroscopy}
We performed spectral analysis for all \chandra\ ACIS data by using
XSPEC v11.3. We also used CIAO's Sherpa for independent  
check. In three of the observations, the X-ray sources are located near 
the aim point and we can use a circular extraction region with radii 
of $0.8-1.3$ arcsec depending on the contamination of nearby
sources. The relatively small extraction radii 
were used because the three X-ray sources that we are interested in are close to each other. 
For the remaining four observations, our targets were off-axis and 
therefore we used an elliptical region to extract the spectra. For the 
background, we used a nearby source free region.
We rebinned the 0.3--7 keV spectra with at least 20 counts per spectral bin, and used 
$\chi^2$ statistics to find the best-fitting parameters. Corresponding
response files were generated using CIAO. 

\begin{figure}[t]
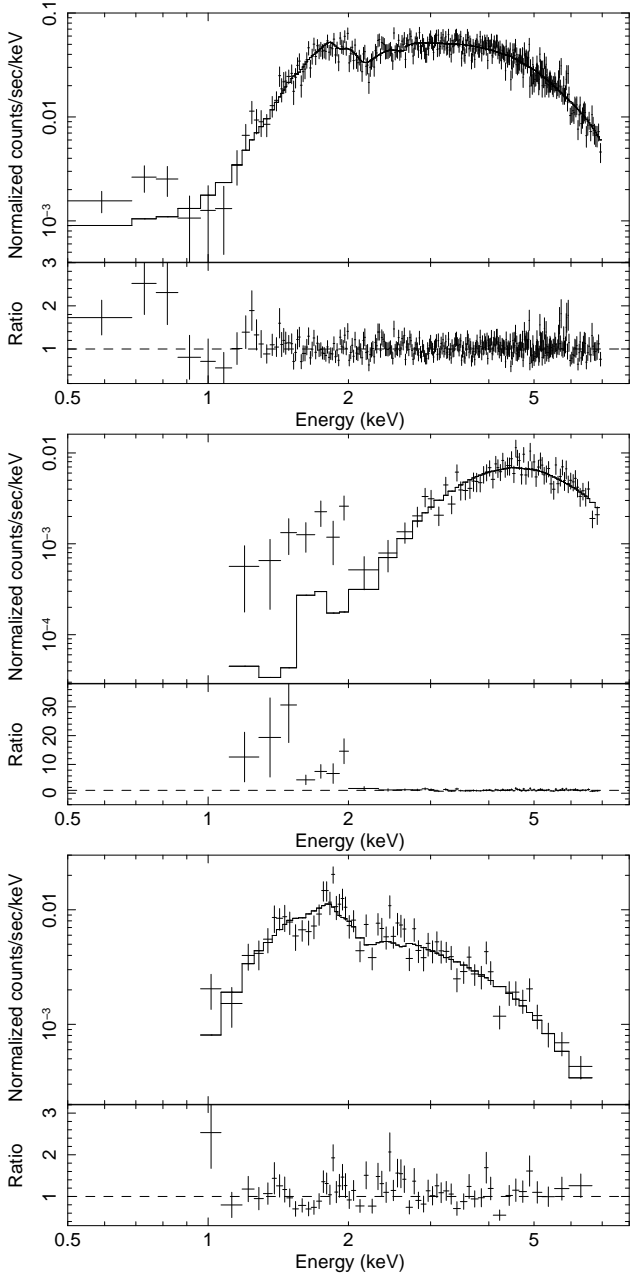

\includegraphics[scale=.35,angle=270]{f2a.eps}\\
\includegraphics[scale=.35,angle=270]{f2b.eps}\\
\includegraphics[scale=.35,angle=270]{f2c.eps}\\
\caption{Power-law spectral fits of J095551.0 (top), J095551.2
  (middle), and J095550.6 (bottom). Spectra are from observation 11. See
  Table 2 for spectral parameters.}
\end{figure}

For the transient (J095551.0), the X-ray spectra can be adequately fitted with an absorbed 
power-law model. The spectral parameters are listed in Table 2. In 
general, the photon index varies between 1.3 and 1.7 while the $N_H$ 
is about $3\times10^{22}$ cm$^{-2}$, consistent with the extinction
measured with near IR observations (Alonso-Herrero et al. 2003). It is worth noting that 
during observation 9, the source suffers mild pile-up. In this case, 
we included a pile-up model in spectral fit yielding a pile-up fraction of 
$\sim 15\%$. For consistency check, we also applied a pile-up model
for other observations. 
Pile-up affects some of the observations (in
particular observations 4, 6, 7 and 8) with a maximum pile-up fraction of
$\sim$ 10\%. Our spectral fit of observation 10 is different comparing
to Kaaret et al. (2006). This is likely due to the diffuse
background and contamination of nearby sources. We used a nearby diffuse emission region as the
background and the fit was performed with the background subtracted
spectrum. We also used a smaller extraction region to reduce the
contamination of nearby sources. In fact, Kaaret et al. (2006)
required an additional very soft
component to fit the spectrum indicating the contribution of diffuse
emission and nearby sources. We also
performed a fit with a bigger extraction region and without background
subtraction. The result is consistent with Kaaret et al. (2006).  
Since five of the observations (4, 6, 7, 8, and 10) are off-axis, the spectra of 
the transient are contaminated by J095551.2 and J095550.6. The
contamination is particularly
serious in observations 4, 6, 7, and 8, and may result the relatively hard spectra of
these observations. To verify the contamination, we extracted combined
spectra of all three sources using observations 11 and 12 for which the
sources are well resolved. While the X-ray flux is dominated by the
transient, the X-ray spectra become significantly harder with a photon index of
$\sim 1$. This indicates that the hard spectra of the three contaminated
observations are likely due to nearby sources. Furthermore, mild
pile-up may also affect the spectra. We therefore do not
include the spectral fits in Table 2.
The best fitting power-law spectrum of observation 11 is shown in Figure 2.

J095551.2 is the next brightest source near the transient. We first
fitted the spectra with an absorbed power-law model and the spectral
parameters are shown in Table 2. Three of the fits are acceptable and
the spectra turn over at about 4 keV suggesting very high
absorption. The best fitted $N_H$ is about $(1-2) \times10^{23}$
cm$^{-2}$ which is an order of magnitude greater than the other two
nearby sources. In addition, all spectra are very hard with
$\Gamma \laeq1$ except for observation 1. 
For observation 11, a soft excess is clearly seen in the
spectrum (Fig. 2) and the fit is much poorer than the others. Indeed,
soft excess is seen in all spectra but with larger uncertainties due
to shorter exposure time or smaller collecting area of ACIS-I below 2 keV.
Soft excess is a common feature of AGN with an ionized absorber. We then
refitted the spectrum (observation 11) with an additional ionized 
absorber ({\it absori} model in XSPEC; Zdziarski et al. 1995). The fit is acceptable with a reduced
$\chi^2$ of 1.06, and the best-fit photon index steepens to
2 with a large absorbing column of $N_H=9\times10^{23}$ cm$^{-2}$
(see Fig. 3); the absorbed 0.5--10 keV flux is $10^{-12}$
ergs cm$^{-2}$ s$^{-1}$.

Apart from observation 11, the X-ray spectra of J095550.6 can be fitted with an 
absorbed power-law 
with $N_H\sim 3\times10^{22}$ cm$^{-2}$ and a photon index of $> 2.7$. 
Observation 11 has the longest exposure time and the X-ray spectrum of J095550.6 is 
clearly more complicated. It cannot be fitted with any single component model. 
Instead, a combination of Raymond-Smith model and power-law model is 
required. In addition, emission lines are clearly seen in the X-ray 
spectrum (Fig. 4). The X-ray spectrum indicates that J095550.6 is either a 
nearby foreground star or a supernova remnant in M82. We will show in 
\S4.3 that J095550.6 is indeed a young supernova remnant in M82.

\begin{figure}[t]
\includegraphics[scale=.35,angle=270]{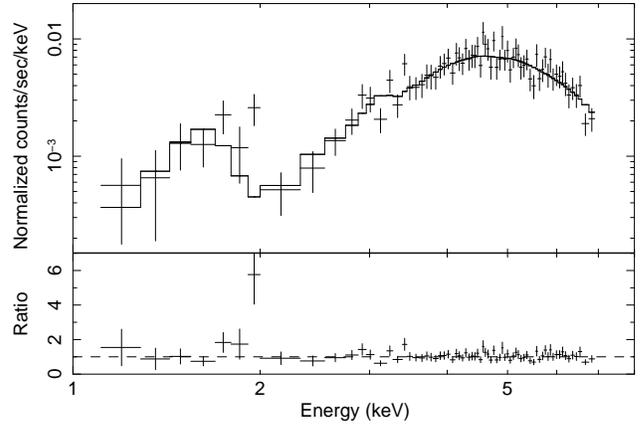}
\caption{\chandra\ spectrum (from observation 11) of J095551.2 with an
  ionized absorber plus power-law model ($N_H=9\times10^{23}$ cm$^{-2}$,
  $\Gamma=2$, ionization parameter $\xi=882$,
  $\chi^2/dof=1.06/69$). We fixed the absorber temperature and Fe
  abundance at $10^5$ K and solar values, respectively
  as the fit was not sensitive to these parameters. 
}
\end{figure}

\begin{figure}
\includegraphics[scale=.35,angle=270]{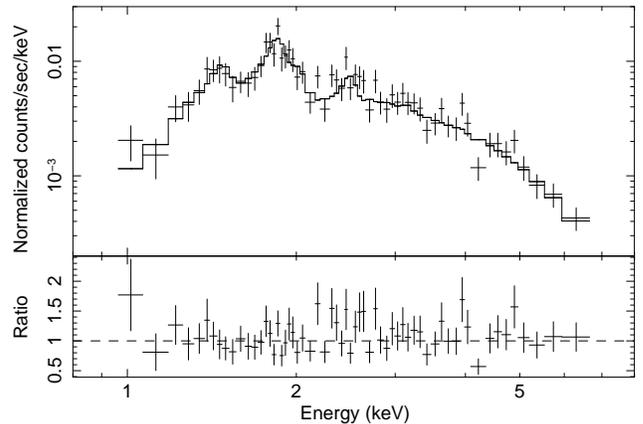}
\caption{\chandra\ spectrum (from observation 11) of J095550.6 with an absorbed
  Raymond-Smith plus power-law
  model ($N_H=2.9\times10^{22}$ cm$^{-2}$, $kT_{RS}=0.9$ keV,
  $\Gamma=2.01$, $\chi^2/dof=1.1/56$). 
}
\end{figure}

\subsection{X-ray Variability}

With the spectral fits, we can estimate the X-ray fluxes of the three
sources and study the long-term variability. We limit our analysis
to \chandra\ data because the three sources as well as M82 X--1 are
not resolved with \einstein, \rosat, and \xmm. The luminous X-ray source, J095551.0, displays strong variability on the timescales of
months (see Fig. 5). In particular, the source was not detected in 1999 September
and reappeared in 1999 October. It was below the detection limit again
in 2000 January and then turned back on in 2000 March. Figure 5 shows the long-term X-ray lightcurve of J095551.0 from 1999
September to 2005 August. For the HRC-I observations and observations
4, 6, 7, and 8, we estimated the
X-ray flux by assuming an absorbed power-law model with
$N_H=3\times10^{22}$ cm$^{-2}$ and a photon index of 1.5. When the
source is active, the X-ray luminosity shows very little variability
at $(7-13)\times10^{39}$ ergs s$^{-1}$. We determined the 90\% upper
limit when the source was undetected; a nearby diffuse emission
region was used as the background since the source is contaminated by
strong diffuse emission. We note that Feng \& Kaaret (2007) reported a
much lower upper limit (without statistical significance) by assuming
the brightest pixel around the source region.

\begin{figure}
\includegraphics[height=3.2in]{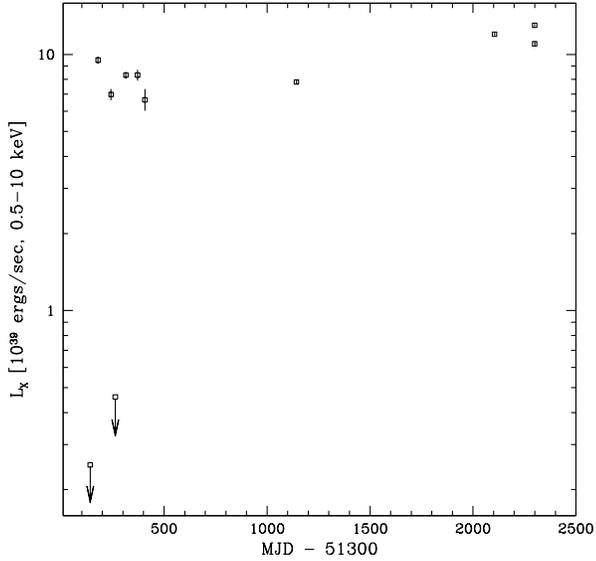}
\caption{Long-term X-ray lightcurve of the ultraluminous X-ray
  transient, J095551.0. The luminosities are determined by spectral
  fits (see Table 2). For the HRC-I observation and observations 4, 6,
  7, and 8, we assume an absorbed
  power-law spectral model with $N_H=3\times10^{22}$ cm$^{-2}$ and
  $\Gamma=1.5$. For non-detections, 90\% upper limits are shown.
}
\end{figure}

\begin{figure*}
\centering
\includegraphics[width=6in]{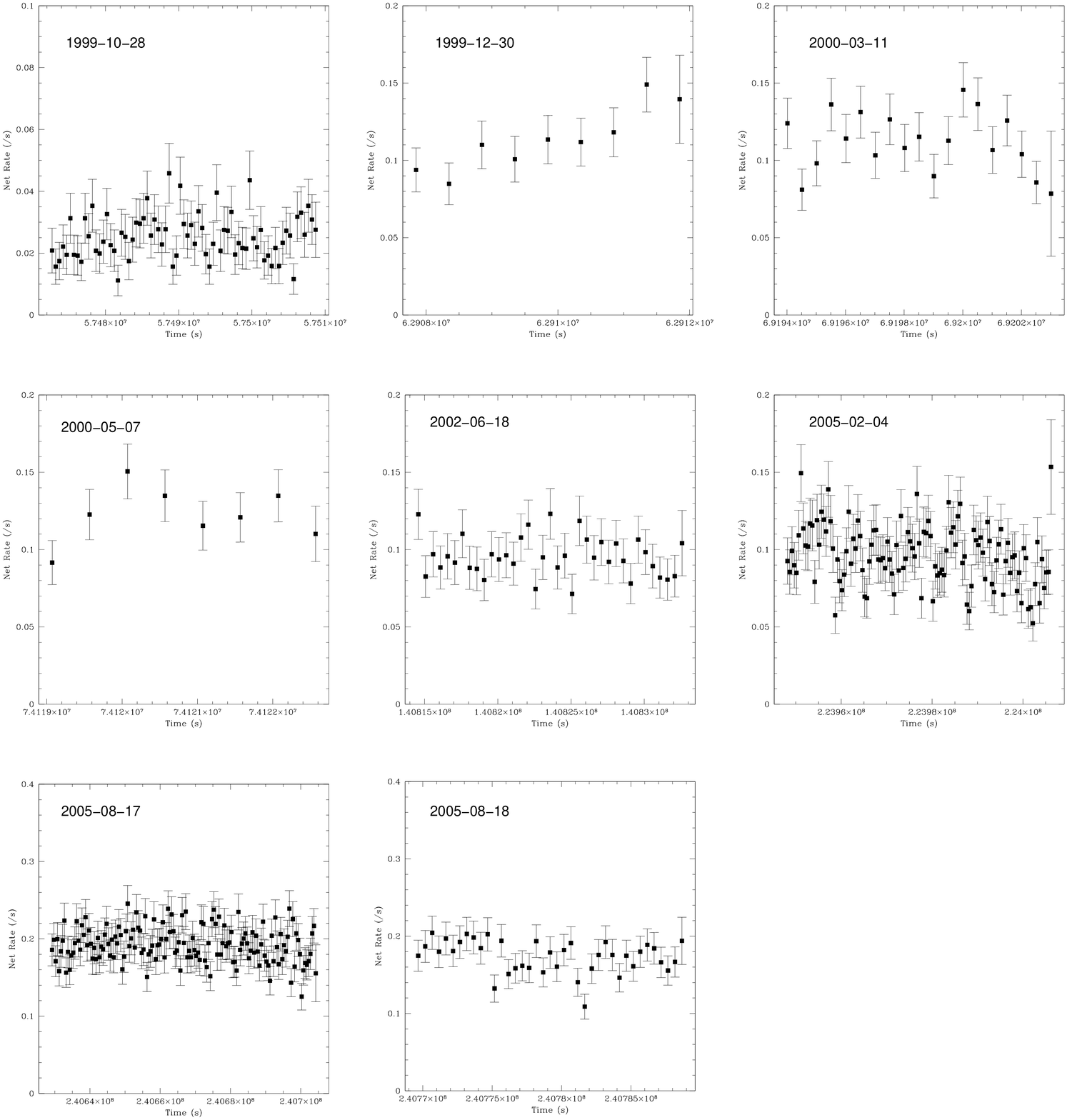}
\caption{\chandra\ short-term lightcurves of the transient, J095551.0,
  when it is active. The time resolution of each plot is 500\,s. We do not show
  observation 8 due to its short exposure time (1.2 ks). We note that
  the apparent difference in the count rate is due to different
  detectors and off-axis angle of the source.
}
\end{figure*}

J095551.2 has a soft spectrum ($\Gamma=2$) during the first observation and it
becomes much harder ($\Gamma \laeq 1$) in subsequent observations. Because of the high $N_H$, the
flux is very sensitive to the photon index. For instance, if we fit
the spectra with an ionized absorber plus power-law model fixing the
spectral parameters except for the normalization as in observation 11, the
luminosities are $\sim 10^{40}$ \lum. Hence, the source does
not show significant variability. 

For J095550.6, the apparent softening during the last observation is
likely an artifact because of the low count rate. We used the same
Raymond-Smith plus power-law model as in observation 11 and the spectrum
can be fitted equally well. The resulting luminosity is about
$2.4\times10^{39}$ \lum, consistent with other
observations. Therefore, the X-ray flux of J095550.6 is consistent
with being constant.

We also study the short-term variability of our targets. 
We extracted the source and background lightcurves from the 0.3--7 keV
event files except that there is no energy filter for HRC-I data. We applied
similar procedure as discussed in \S\,3.2 to define the source and
background regions of our targets. All three sources do not show
significant variability on timescale of hours. We show the short-term lightcurves
of the transient in Figure 6.

\subsection{Radial Profile}
We investigated the spatial extent of the ULXs using observation 11 for
which the sources have the highest number of counts and are well resolved. Soft
(0.3--3 keV) and hard (3--7 keV) band counts were extracted from
energy filtered images. We also modelled the point spread
functions (PSF) of each source
using \chandra\ Ray Tracer (ChaRT) and compared them with the measured
radial profile of the three sources.
Counts from these images were extracted in identical manner
and normalized by the innermost nuclear annulus. The resulting radial
profiles are shown in Figure 7.

The surface brightness distributions of all sources but
J095551.2 in the hard and soft bands are very similar. They are
centrally peaked and lie above background level out
to radius $\approx2''$ with the exception of J095551.2 for which the
contamination from background is significant in the soft band. The radial profiles
were compared with the PSF models; we do not find significant
difference based on a Kolmogorov-Smirnov (K-S) test for all sources
but J095551.2.

\begin{figure}[t]
\center{
\includegraphics[width=2.6in]{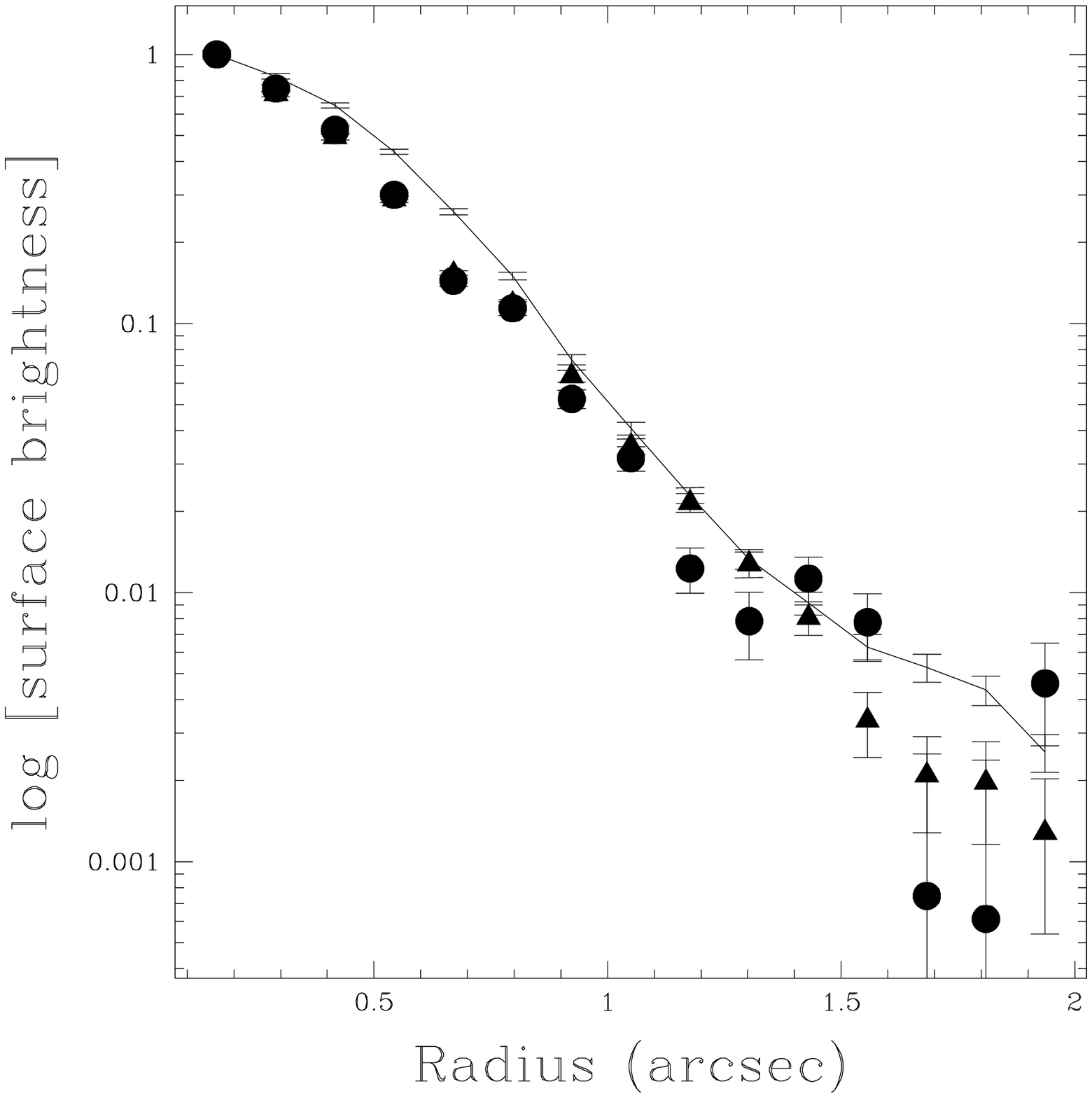}\\
\includegraphics[width=2.6in]{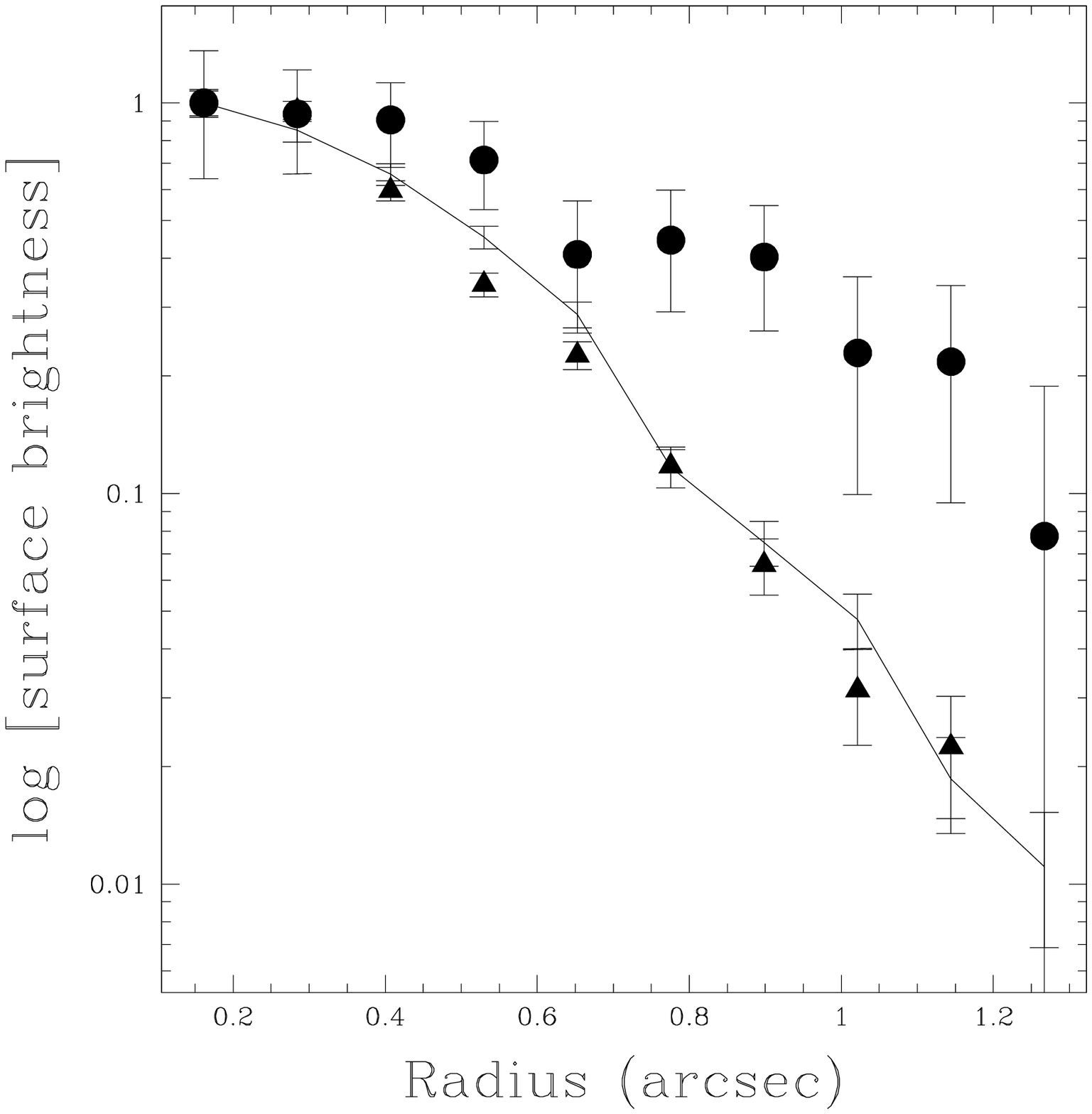}\\
\includegraphics[width=2.6in]{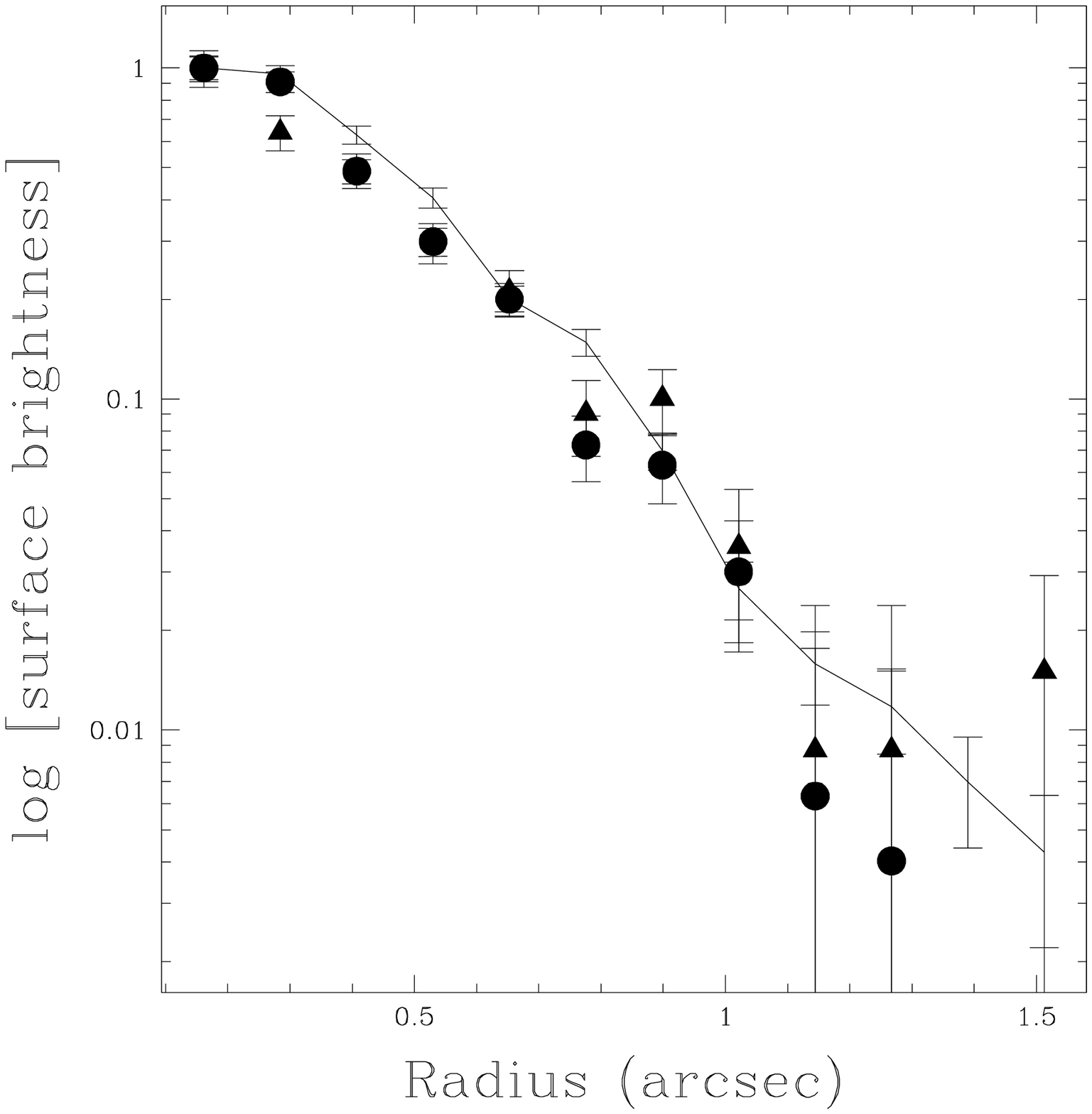}\\
}
\caption{The soft (0.3--3 keV; solid points) and hard (3--7 keV;
  triangles) band radial profile of J095551.0 (top), J095551.2
  (middle), and J095550.6 (bottom), compared to a \chandra\ PSF model
  (solid curve), from observation 11. 
}
\end{figure}

\subsection{Near IR Imaging}
In order to compare the \chandra\ and \hst\ NICMOS images, we first
aligned the two images. For the X-ray image, we used the HRC-I
observation taken on 1999 October 28 (observation 3) as the reference
frame because it has a wide field-of-view and moderate
exposure. Furthermore, all three targets were active during this
observation.
We used CIAO tool {\it wavdetect} to detect X-ray sources in the HRC-I image.
We compared the X-ray source list with the 2MASS catalog, and looked
for coincidence of bright and isolated stellar 
objects. We found one star (2MASS\,09551494+6936143) that is $< 1''$ from
the corresponding HRC-I position.
From the ACIS-I observation (observation 9), the X-ray colors of the X-ray emitting star indicate that it
has a very soft X-ray spectrum (84\% of the source counts come from $<
1$ keV with no counts above 2 keV), consistent with a very soft X-ray source
(Di\,Stefano \& Kong 2004).
The X-ray radiation is therefore likely due to the coronal
emission from a foreground star. The star has a $R$ magnitude of 10.1
(Monet et al. 2003). We calculated the X-ray to optical flux ratio as
$\log(f_X/f_R)=\log f_X + 5.67 + 0.4 R$ (Hornschemeier et al. 2001).
With a count rate of $9.4\times10^{-4}$ c/s in the ACIS-I detector and assuming a
Raymond-Smith model with
$kT_{RS}=0.3$ keV and $N_H=4\times10^{20}$ cm$^{-2}$ (the Galactic
value toward the direction of M82), the 0.3--10 keV flux is
$10^{-14}$\flux and the
corresponding $f_X/f_R$ is $5.1\times10^{-5}$, consistent with a foreground star.

Based on the 2MASS counterpart, the boresight
correction that needs to be applied to the X-ray source positions is
$0.87\pm0.56$ arcsec in R.A. and $0.73\pm0.27$ arcsec in decl.; the uncertainties are
a quadratic sum of the errors on the X-ray and 2MASS
positions. To study the IR environment of the X-ray sources, we plot
on the NICMOS image (Figure 8) the corrected X-ray positions with error circles
given by the quadratic sum of the positional uncertainty for the
X-ray source ($0.032''$ for the transient and $0.063''$ for the other
two sources), the uncertainty in the optical astrometry (2MASS to ACS
astrometry and ACS to NICMOS astrometry; $0.2''$), and the
uncertainty in the X-ray boresight correction ($0.62''$). In the figure, we also plot the locations of
several known super-star clusters and M82 X--1. As expected, this
area shows many star forming regions with the presence of
super-star clusters. The three luminous X-ray sources as well as M82
X--1 are located near star clusters. In particular, a near IR source
is at the center of the error circle of J95551.0. The source is
marginally resolved in the F160W image with a half-light radius of 0.6 pc by fitting with
a King model (N. McCrady, private communication). At that size the
cluster likely has a mass of $< 10^5 M_\odot$. 
J095551.2, however, does not seem to have
any counterpart. While there is no obvious counterpart within the
X-ray error circle of J095550.6, some unresolved IR emission is
seen. Furthermore, a super-star cluster, MGG--8, is just outside the
$1\sigma$ X-ray error circle locating 1 arcsec from the source. It is also worth noting
that M82 X--1 is located just outside the super-star cluster, MGG--11. 

\begin{figure*}
\centering
\includegraphics[width=6.5in]{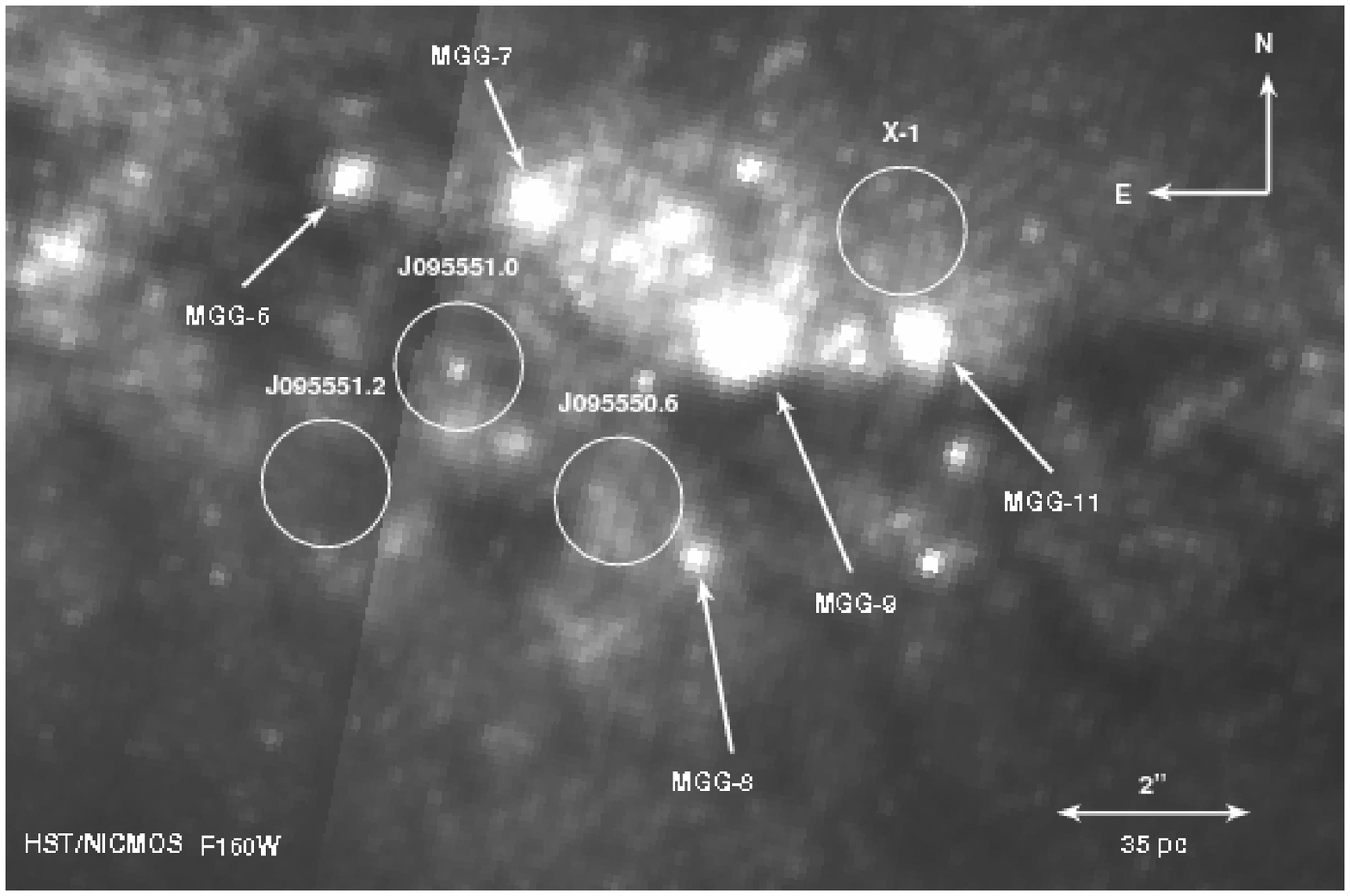}
\caption{\hst/NICMOS F160W image of the region around the X-ray
  sources analyzed in this paper (J095551.0, J095551.2, and J095550.6). The $1\sigma$ \chandra\ error
  circles (0.66 arcsec) are shown. We also label the locations of
  known super-star clusters and M82 X--1.
}
\end{figure*}

\section{Discussion}
\subsection{J095551.0: An Ultraluminous X-ray Transient}

The most intriguing behavior of J095551.0 is the X-ray
variability. The source varies from below the detection limit,
$\sim 2.5\times10^{38}$ ergs s$^{-1}$ in the 0.5--10 keV band, to
$\sim 10^{40}$
ergs s$^{-1}$ on timescales between observations of $\sim 2$
months. Furthermore, the source shows recurrent outbursts. Recurrent
ultraluminous transients are not common in nearby galaxies. By
comparing \rosat\ and \xmm\ observations, Winter et al. (2006) found
that most of the ULXs are persistent sources with less than a factor
of 3 in flux variation over the timescale from \rosat\ to
\xmm. Nevertheless, ultraluminous X-ray transients have been found in 
NGC 3628 (Strickland et al. 2001), M74 (Soria \& Kong 2002), NGC 300 (Kong \&
Di\,Stefano 2003), NGC 253 (Bauer \& Pietsch 2005), and M101 (Kong et
al. 2004; Kong \& Di\,Stefano 2005). Two of these sources (NGC 300 and
M101) are ultraluminous supersoft sources with $kT \laeq 0.1$ keV. The
ones in NGC 3628 and M74 are typical power-law sources with
$\Gamma\sim 2$ while
the ULX in NGC 253 can be described with a bremsstrahlung model with
$kT=2.2$ keV. J095551.0, however, has a much harder spectrum. These
sources also show diverse luminosity range. The sources in NGC 253,
M74, and NGC 300 only reach $L_X \gaeq 10^{39}$ ergs s$^{-1}$ suggesting that
an IMBH is not necessary. On the other hand,
the sources in NGC 3628 and M101 can have 0.3--8 keV luminosities of
$\sim 5\times10^{40}$ ergs s$^{-1}$ with bolometric luminosities
approaching $10^{41}$ ergs s$^{-1}$. While we cannot totally rule out
a stellar-mass black hole model, a black hole of intermediate mass is
certainly an attractive scenario.

A factor of $> 50$ in flux variation indicates that J095551.0 is a compact source while
recurrent outbursts and hard X-ray spectra can rule out the possibility that the source is a
young supernova remnant. It is also unlikely to be a background AGN since
AGN normally varies by a factor of $< 10$ on timescales of days to months. However, transient AGNs are not
unusual (e.g., Komossa et al. 2004; Grupe et al. 2004) and they belong to a class of AGNs called 
narrow-line Seyfert 1 galaxies. These galaxies have
X-ray spectra much softer ($\Gamma > 2.5$; e.g., Boller et
al. 1996; Grupe et al. 2004) than J095551.0. The more likely scenario is
that J095551.0 is a binary system with a black hole accretor. The high
X-ray luminosity ($L_X\approx10^{40}$ ergs s$^{-1}$) when it is active indicates that it
may be an ULX with an IMBH. Assuming the
emission is isotropic, the X-ray luminosity implies that the compact
object is a $\sim 100 M_\odot$ black hole. Many ULXs have a thermal
component with a temperature of $\approx 0.1$ keV which is interpreted
as evidence of IMBHs with masses of $\sim 100-1000
M_\odot$ (Miller et al. 2004). J095551.0, however, does not have any
soft excess and the X-ray spectra are well fitted with an absorbed power-law
model with photon index $\Gamma =1.3-1.7$. This resembles to the low/hard
state of Galactic black hole X-ray binaries (McClintock \& Remillard
2006). 
It is therefore possible that the source can be explained in the framework of the
advection-dominated accretion flow (ADAF) model. 

More recently, Yuan
et al. (2007) apply an ADAF model to describe the X-ray emission of
M82 X--1 and argue that the accreting compact object is an IMBH.
During the low/hard state, the ADAF model predicts that the X-ray
luminosity is about $< 1\%-10\%$ of
the Eddington luminosity. If J095551.0 is accreting at a rate similar
to the hard-state of Galactic black hole X-ray binaries, this would imply a black hole mass
of $\sim 1000-10^4 M_\odot$. 
The estimate should be treated with caution because we assume that the
X-ray spectra of the transient ULX are similar to that of Galactic
black holes in the low/hard state. 
Indeed, pure power-law spectral model for
ULXs is not uncommon; \chandra\ and \xmm\ observations have revealed
hard-state ULXs in several nearby galaxies (see e.g., Roberts et
al. 2004; Winter et al. 2006). Hard-state ULXs may be good candidates
to IMBHs. If the accreting object is instead a
stellar-mass black hole in the hard state, the X-ray emission must be
anisotropic in order to produce such a high X-ray
luminosity. However, the inner accretion disk of hard-state Galactic
black hole X-ray binaries are truncated at large distances from the
black hole and this may be a problem for the thick-disk plus central funnels
anisotropic radiation model (King et al. 2001). Furthermore, the lack
of short time variability of J095551.0 argues against the relativistic beaming
model since this would require a very stable jet.
It is worth noting that the observed photon index of J095551.0 is sometime
harder than the typical hard-state value of $1.5 < \Gamma < 2.1$ for
Galactic X-ray binaries (McClintock \& Remillard 2006). It is
therefore not clear if we can directly compare with Galactic black
hole X-ray binaries in the hard state. Alternatively, it may be a
unique state that the ULX is a stellar-mass black hole accreting at very high rate.

In addition to the long-term timing variability, the X-ray spectra
also vary. Excluding those observations taken with off-axis pointings,
the photon index is consistent with 1.5--1.7 except for the last
observation (see Table 2). In the last observation, the photon index
becomes much harder with $\Gamma=1.27\pm0.18$. Moreover, the spectral
change is quite dramatic. The observation taken one day earlier has a
photon index of $1.52\pm0.10$ while the X-ray luminosity does not
change significantly. On the other hand, the nearby source, J095551.2,
does not show this dramatic change in the spectra. We also check the
spectra of a few bright sources in the field for the last two
observations; none of the sources displays this kind of spectral
variability. We therefore can conclude that the spectral hardening is
real. Such an X-ray spectrum is unusual for Galactic black hole X-ray
binaries. The only exception is the fast X-ray nova, SAX\,J1819.3--2525 whose spectrum is
extraordinary hard with $\Gamma=0.6-0.9$ during a flaring state (Markwardt et al. 1999;
McClintock \& Remillard 2006). The last observation of J095551.0 is
similar to SAX\,J1819.3--2525 but we note that the luminosity is
indeed slightly lower than that measured one day earlier and the
short-term light curve does not display strong variability.

Near IR and radio observations may provide additional clues about
the nature of J095551.0. From the NICMOS image (Fig 8), a star cluster is at the center of the X-ray
error circle of J095551.0, suggesting that the source is associated
with the cluster. Indeed, young star clusters are ideal places to
produce IMBHs via the collapse of very massive
stars through runaway stellar collisions (Portegies Zwart et al. 2004). The coincidence of J095551.0 and
a star cluster strongly suggests that the ULX is produced in the
cluster and is consistent with a black hole of intermediate mass.
Radio emission is also
detected within the X-ray error circle (K\"ording et al. 2005; Kaaret
et al. 2006). The radio source, known as 42.21+59.0, has been detected
several times (e.g., McDonald et
al. 2002). The flat spectrum (between 5 and 15 GHz) and extended size
(4.9 pc) suggest that it is a giant \ion{H}{2} region with 94 O5 stars
(McDonald et al. 2002). 

\subsection{J095551.2: A Highly Absorbed X-ray Source}

The X-ray spectrum of J095551.2 is very different compared to that of other
nearby sources. The source has an unusually high $N_H$ ($> 10^{23}$
cm$^{-2}$) while the two nearby sources as well as M82 X--1 (Kaaret et
al. 2006) have a $N_H$ of $\sim 3\times10^{22}$ cm$^{-1}$, consistent
with the extinction measured by IR observations. Furthermore,
J095551.2 has a very flat spectrum ($\Gamma \laeq 1$) with
soft excess below 2 keV.  The high absorption column density  may indicate that
the source is a background AGN. AGN normally has a power-law spectral
model with $\Gamma \sim1.7-2$ (e.g., Page et al. 2006) while the spectrum of J095551.2
is much harder. A hard spectral index of AGN may be due to the
presence of a reflection component and/or a complex absorber (Cappi et
al. 2006). In \S3.1, we refitted the spectrum (observation 11) with an
ionized absorber plus power-law model yielding an acceptable fit. The
photon index steepens to 2 and the 0.5--10 keV unabsorbed flux is
significantly higher. The other two spectra (observations 9 and 12) can
also be well fitted with an ionized absorbed plus power-law model with
the spectral parameters fixed at the values determined in observation 11
except for the normalization. However, due to the low count rate of
these two observations, it is not clear if the difference is real. On
the other hand, the spectrum of the first observation 
might be different. The spectral parameters of observation 11 cannot fit
the spectrum of the first observation. If we only fit all the spectra in the range of 2.5-7 keV
where soft excess is not crucial, the spectrum of the first
observation is much softer ($\Gamma=2.9$) than the others ($\Gamma
\laeq1$). Hence, the spectral change during the first observation
might be real. Spectral change is also seen in some narrow-line
Seyfert 1 galaxies such as
NGC 4051 (Guainazzi et al. 1996; Ponti et al. 2006; see Leighly 1999
for a review). 
Therefore, J095551.2 is likely to be a highly absorbed background
AGN. It is worth nothing that there is a \ion{H}{2} region within the
X-ray error circle (42.56+580; McDonald et al. 2002).

\subsection{J095550.6: A Young X-ray Supernova Remnant}

J095550.6 is unique because the X-ray spectrum cannot be fitted with
simple spectral models. Instead, a combination of a Raymond-Smith
model and a power-law model is required to describe the
spectrum (see Fig. 4). Given the high $N_H$ that is consistent with the extinction
of nearby region of M82, J095550.6 is unlikely to be a foreground
star. The X-ray spectrum is not typical for X-ray binaries. We can also rule out a background AGN due to its unusual
spectrum. The remaining possibility is that it is an X-ray luminous supernova
remnant. In particular, it
is evident that the spectrum (Fig. 4) is dominated by broad emission of
Mg XII lines at 1.4 keV, Si K shell lines at 1.8 keV, and S K shell
lines at 2.5 keV. Strong emission lines are also seen in some luminous
X-ray emitting supernova such as SN 1978K (Schlegel et
al. 2004). Indeed, SN 1978K is also the first known supernovae with X-ray
luminosity above $10^{39}$ ergs s$^{-1}$. For direct comparison with
SN 1978K, we refit the spectrum of J095550.6 with an absorbed MEKAL + power-law
model. The spectral parameters are not sensitive to the model, with a
best-fit temperature of 0.9 keV. This is slightly hotter than SN 1978K
($\sim 0.7$ keV; Schlegel et al. 2004). The 0.5--10 keV luminosity of
J095550.6 is $2.2\times10^{39}$ ergs s$^{-1}$. 

Within the X-ray error circle,
there is a strong radio source known as 41.95+575 (McDonald et
al. 2001,2002; K\"ording et al. 2005). 41.95+575 is the brightest and most
compact radio source in M82 and has been detected since 1965 (Muxlow
et al. 2005). High resolution (3 mas) VLBI imaging shows that the source has a
double-lobed structure (McDonald et al. 2001). The separation of the
two brightest components is 22.4 mas in 2001 (Muxlow et
al. 2005). Furthermore, multiple-epoch VLBI observations show that the
separation is increasing at a rate of 0.24 mas yr$^{-1}$. The radio
flux also varied over the last 30 years. It has decreased in flux
density at a rate of $\sim 8.8\%$ per year. An age of around 100 years is
estimated (Muxlow et al. 2005). The simplest explanation of the nature
of J095550.6 is that it is a supernova event taking place within a
high density molecular cloud (McDonald et al. 2001). We can also
estimate the age of the remnant using the X-ray spectral
fit. Following Kong et al. (2002), assuming J095550.6 is in the
adiabatic expansion phase, the shock temperature can be written as
$T_s= (0.18~\mbox{keV})(R/t_3)^2$, where $R$ and $t_3$ are the radius (in
units of parsecs) and age (in units of 1000 yr),
respectively. Adopting a radius of 11 mas ($=0.19$ pc), and $T_s=0.9$
keV, we obtain $t\sim100$ yr which is consistent with radio
observations. It is worth noting that long-term radio observations have suggested that 41.95+575
may be a radio afterglow of a 100 year old gamma-ray burst event
(Muxlow et al. 2005).

\section{Conclusions}
We have used archival \chandra\ and \hst/NICMOS data to study the
physical nature of three ULXs near the center of M82. We found a
recurrent transient ULX, J095551.0 from the \chandra\ data. During its
active state, the X-ray luminosity is about $7\times10^{39} -
1.3\times10^{40}$\lum\ and it was turned off twice in 1999 and 2000 indicating
a factor of $> 50$ variability. This also rules out the possibility that it
is a supernova remnant. The X-ray spectra can be fitted with a
power-law model with photon index $\Gamma=1.3-1.7$ which is similar to
Galactic black hole X-ray binaries in the low/hard state. We suggest
that the X-ray emission might be explained in the framework of the
ADAF model implying a black hole mass of $\sim 10^3-10^4
M_\odot$. However, we cannot totally rule out that the source is 
in a unique spectral/luminosity
state. In particular, spectral hardening was seen in one of the
observations. We also examined near IR images taken with \hst/NICMOS. We
found a star cluster at the center of the X-ray error circle
suggesting that the source is associated with the cluster.

With an unusually high column
density $N_H>10^{23} cm^{-2}$ and a rather flat X-ray spectrum ($\Gamma
\laeq 1$) with an ionized absorber, it is suggested that the source J095551.2 is likely to be a
background AGN. The source also shows spectral change similar to some
narrow-line Seyfert 1 galaxies. In addition, there is a \ion{H}{2} region known as
42.56+580 associated with the X-ray source. 

The source J095550.6 shows an unusual spectrum that cannot be fitted with a simple power-law
model. Instead, the spectrum can be fitted with a Raymond-Smith model, accompanying with broad
emissions of Mg, Si, and S, indicating that it is a supernovae remnant
with an age of $\sim 100$
years. Furthermore, a radio source known as 41.95+575 is found
associated with J095550.6 within the X-ray error circle. The long-term
radio observations reveal that 41.95+575 may be a radio afterglow of
a 100-year old gamma-ray burst event. 

\begin{acknowledgements}
Part of this work was carried out at the National Tsing Hua
University, Taiwan, and we thank Hsiang-Kuang Chang for warm hospitality.  
We also thank Nate McCrady for providing the structural parameters of the
star cluster associated with the transient ULX.
The \hst\ data presented in this paper were
obtained from the Multimission Archive at the Space Telescope Science Institute
(MAST). STScI is operated by the Association of Universities for
Research in Astronomy, Inc., under NASA contract NAS5-26555.
\end{acknowledgements}

{\it Facilities:} \facility{CXO (ACIS, HRC)}, \facility{HST (NICMOS, ACS/WFC)}

\end{document}